\documentclass[aps,prb,twocolumn,superscriptaddress,showpacs]{revtex4}
\usepackage{graphicx}
\usepackage{calc}
\usepackage{bm}

\begin{document}

\title{Topological defects in the edge state structure in a
  bilayer electron system.}

\author{E.V.~Deviatov}
\email[Corresponding author. E-mail:]{dev@issp.ac.ru}
 \affiliation{Institute of Solid State
Physics RAS, Chernogolovka, Moscow District 142432, Russia}

\author{V.T.~Dolgopolov}
\affiliation{Institute of Solid State Physics RAS, Chernogolovka,
Moscow District 142432, Russia}

\author{A.~W\"urtz}
\affiliation{Laboratorium f\"ur Festk\"orperphysik, Universit\"at
Duisburg-Essen, Lotharstr. 1, D-47048 Duisburg, Germany}

\author{A.~Lorke}
\affiliation{Laboratorium f\"ur Festk\"orperphysik, Universit\"at
Duisburg-Essen, Lotharstr. 1, D-47048 Duisburg, Germany}

\author{A.~Wixforth}
\affiliation{Institut f\"ur Physik, Universitat Augsburg,
Universitatsstrasse, 1 D-86135 Augsburg,  Germany}

\author{W.~Wegscheider}
\address{Institut f\"ur Angewandte und Experimentelle Physik,
Universitat Regensburg, 93040 Regensburg,
 Germany}

\author{K.L.~Campman}
\affiliation{Materials Department and Center for Quantized Electronic
Structures, University of California, Santa Barbara, California
93106, USA}

\author{A.C.~Gossard}
\affiliation{Materials Department and Center for Quantized
Electronic Structures, University of California, Santa Barbara,
California 93106, USA}

\date{\today}

\begin{abstract}
We experimentally demonstrate, for the first time,  formation of
point-like topological defects in the edge state structure in the
quantum Hall effect regime. By using of a selective population
technique, we investigate  equilibration processes between the
edge states in bilayer electron structures with a high tunnelling
rate between layers. Unexpected flattening of the $I-V$ curves in
perpendicular  magnetic field at a specific filling factor
combination and the recovery of the conventional nonlinear $I-V$
characteristics in tilted fields give a strong evidence for the
existence  of topological defects.
\end{abstract}


\maketitle

Since its introduction by Halperin~\cite{halperin}, the concept of
the edge states (ES) was found to be useful in describing
transport phenomena in two-dimentional (2D) systems for both
sharp~\cite{buttiker} and smooth~\cite{shklovsky} edge potential
profile. ES are arising in a quantizing magnetic field at the
intersections of the Fermi level with distinct Landau levels,
which are bent up by the edge potential.

Much attention was paid to substantiate the ES picture
experimentally by investigating electron transport along ES, as
well as between them (for a review, see Ref.~\onlinecite{haug}).
Different imaging techniques~\cite{shashkin,klitzing} were used to
demonstrate the existence of compressible and incompressible
strips of the electron liquid at the sample edge, which is widely
accepted nowadays.

However, there is another class of phenomena in ES that has not
been investigated yet. It is the formation of so-called
topological defects, which have been predicted to occur, e.g.,
when two ES with different spins locally switch their positions
and thus cross each other at several (at least two)
points~\cite{demsley,bauer}. Possible mechanism for such a
crossover was proposed theoretically for a non-equilibrium
occupation of spin-split ES, when the chemical-potential
difference is of the order of the spin splitting~\cite{bauer}.
This situation can not be achieved in usual experiments performed
in the Hall-bar geometry~\cite{mueller}. In recent experiments in
quasi-Corbino geometry~\cite{alida,rdiff} with direct biasing of
the ES, no sign of ES crossing was observed, even when the ES
imbalances exceeded the cyclotron gap~\cite{relax}. This may be
due to the fact that even at the crossing points, the electron
spin flips, which are necessary for the inter-ES scattering, can
be strongly suppressed~\cite{mueller}.

More pronounced features may be expected in the transport
properties of bilayer systems, where not only spin, but also
isospin, related to the layer index~\cite{zheng}, is involved. The
energies of the isospin states can be tuned in situ by application
of a suitable gate voltage, which,  as we will show below, gives
direct control over the presence of the topological defects.

Here, we experimentally study topological defects (namely, the
edge state crossings) in a bilayer electron system. We use a
selective population technique to investigate equilibration
processes between different edge states. By adjusting the gate
voltage and the magnetic field, we force the bilayer system to be
in different isospin states in the gated and ungated regions, that
leads to a crossing of the ES, which we detect in the transport
characteristics.

\begin{figure}
\includegraphics*[width=0.95\columnwidth]{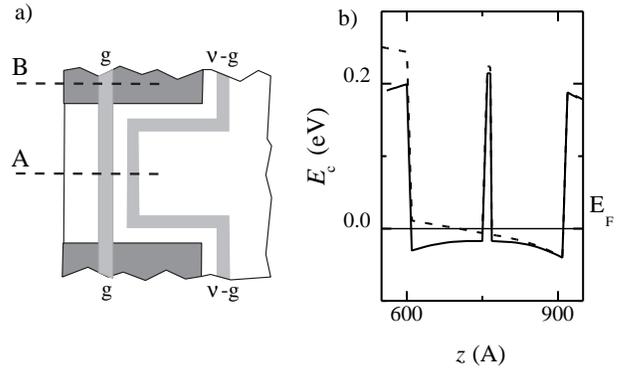}%
\caption{ a) Schematic diagram of the gate-gap region in the
pseudo-Corbino geometry.  The shaded area represents the
Schottky-gate. In a quantizing magnetic field at total filling
factor $\nu$, one set of the edge states (the number is equal to
the filling factor under the gate, $g$; $g<\nu$) is propagating
under the gate along the etched edge of the sample, while the
other edge states (their  number is $\nu-g$) are going along the
gate edge. In the gate-gap region, both sets of the edge states
are running in parallel. b) Quantum well subband diagram at zero
gate voltage (solid line) and at smaller electron concentration,
near bilayer onset (dashed line)\label{sample}}
\end{figure}

Measurements are performed in a quasi-Corbino sample geometry, in
which a top gate of special shape is used to independently contact
ES and bring them into interaction over a distance which is much
smaller than the equilibration length~\cite{alida}. In Fig. 1 (a),
only the interaction region of the sample is shown. Voltage V ,
applied between contacts that are connected to the inner and outer
ES (not shown in Fig.~\ref{sample}), produces an energy shift $eV
(e < 0)$ between them. This allows to directly investigate the
charge transfer between ES in the interaction region (gate-gap) at
imbalances even higher than the spectral gaps.

In gated bilayer systems, such as the one with quasi-Corbino
geometry used here, the gate bias does not only influence the
lateral, but also the vertical distribution of charge. At zero
gate voltage, the bilayer systems studied here are slightly
imbalanced (i.e. they have different electron concentrations in
two parts of the well). A negative top-gate bias increases this
initial imbalance, see Fig. 1 (b). The electric field mostly
affects the electron density   in the top (closest to the gate)
layer of the well, which will deplete at some bias $V_{on}$ (the
bilayer onset voltage). At lower gate voltages, only the back part
of the well is filled, so that the electron system becomes a
single-layer. Thus, even for a sample where the edge states in the
gate gap originate from the bilayer spectrum, the channels under
the gate may originate from either a bi- or a single-layer
spectrum. By using samples with different $V_{on}$ it is therefore
possible to realize different regimes under the gate, even at the
same filling factor. For a single-layer system in a quantizing
magnetic field, the energy spectrum is the usual Landau ladder of
energy levels. In the bilayer regime, the situation is more
sophisticated~\cite{dqw,tilt,davies,japan}. In imbalanced bilayer
systems, the electrons belong to a particular electron sheet,
which makes it necessary  to introduce a new quantum number:  the
isospin, or layer index, equal to  $\pm 1$ for top/bottom layer,
correspondingly). In sufficiently high magnetic fields
(corresponding to low filling factors $\nu=$1 and 2), and/or in
balanced electron system, the  electrons have non-zero wave
functions in both layers and, consequently, are in a mixed isospin
state. The energy spectrum is then a single Landau ladder with an
additional energy gap: the symmetric-antisymmetric splitting
$\Delta_{SAS}$.

Another way to couple both layers is through application of an
in-plane magnetic field component. Not only does it increase the
Zeeman splitting with respect to other energy scales, but it also
forces common subband formation~\cite{tilt} even at high filling
factors $\nu>2$.

\begin{figure}[t]
\includegraphics*[width= 0.5\columnwidth]{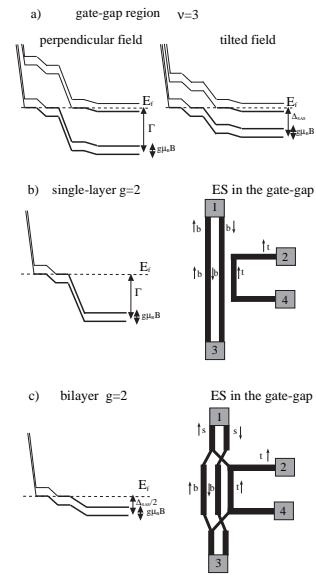}%
\caption{ Energy subband diagram of the sample edge  for the
filling factors $\nu=3$ and $g=2$ with no voltage applied between
the ES . a) In the gate-gap region  in perpendicular and tilted
magnetic fields. b) Under the gate in a single-layer regime
(sample B) and the resulting ES structure in the gate-gap c) Under
the gate in a bilayer regime (sample A) and the resulting ES
structure in the gate-gap. Symbols $t, b, s, \uparrow, \downarrow$
indicates the isospin (top, bottom, symmetric) and the spin (up
and down) states of the ES. Ohmic contacts, connected to different
ES, are denoted by gray bars. \label{ES}}
\end{figure}

In the present paper we investigate the samples  at filling factor
 $\nu = 3$ in the gate-gap. In perpendicular magnetic
field, ES originate from two independent Landau ladders (see
Fig.~\ref{ES} a). The separation between the two outer spin-split
ES and the single inner one is given by the difference $\Gamma$ of
the subband energies of the two electron layers.

Let us first consider the case when the electron system under the
gate is in a single-layer regime for filling factor g = 2. Two
spin-split energy levels in the bottom electron layer form two ES,
which are directly connected to the two outer spin-split ES in the
gate-gap, because  the same spin and isospin indexes are the same
(see Fig.~\ref{ES} b).  Applying a bias voltage to the ohmic
contacts 1 and 2  in Fig.~\ref{ES} produces an energy shift of the
two outer ES in the gate-gap with respect to the inner one.
Positive voltage $V>0$ flattens the potential relief between the
ES~\cite{alida,rdiff} because of $eV<0$. Significant current
starts to flow only after complete flattening of the subband
bottom, i.e. at $V_{th}^+ \sim\Gamma$, so this branch of the $I-V$
curve should contain a clear-defined threshold. Applying a
negative voltage $V<0$ produces only a tunnel current between ES,
leading to a strongly non-linear $I-V$ branch.

Let us now turn to the case where the gated region is in the
bilayer regime at $g=2$. Then the one to- one correspondence
between the edge states in the ungated and gated parts is no
longer possible and a new ES topology will develop, as shown in
Fig.~\ref{ES} (c), right. Under the gate, two energy levels are
filled. They both correspond to the lowest (symmetric) subband and
are separated by the spin gap, see Fig.~\ref{ES} (c), left.
 Electrons in the corresponding ES
belong to both electron layers simultaneously. Roughly speaking,
there is $50\%$  probability to find an electron in either layer.
When a charge is injected into the gate-gap region, both spin and
isospin should be conserved. Isospin conservation means that
approximately half of electrons are injected into the top electron
layer and the others are injected into the bottom layer. Taking
into account the spin conservation, we conclude that the electrons
from the outer ES under the gate will be injected into both the
outermost and the innermost ES in the gate-gap, so they will share
the same chemical potential. Thus, the ES can cross each other in
the corners of the gate-gap, as sketched in Fig.~\ref{ES} (c),
right. Two ES with the same spin in the gate-gap will be
full-equilibrated, leading to linear   $I-V$ curve.

The described picture can easily be destroyed by an in-plane
magnetic field component. As mentioned above, in tilted magnetic
fields the states in the gate-gap region will also be delocalized
between both layers. As shown in Fig.~\ref{ES} (a), right, two
spin-split ES will originate from the lowest, (symmetric) quantum
well state, while the third ES will be associated with the
antisymmetric state, separated by $\Delta_{SAS}$. Because the two
outermost edge states now have the same isospin in both gated and
ungated regions of the sample, the ES topology will resemble the
one shown in Fig.~\ref{ES} (b), left. Therefore, there will be no
ES crossing   and the I-V curves should become strongly non-linear
again, as in the single-layer case.  The height of the step on
positive branch  of I-V curve will now be given by $\Delta_{SAS}$.

From the picture above described  we expect that ES crossing can
be realized in samples where the gated and ungated parts are in a
different isospin configuration. The equilibration caused by the
crossing should lead to a linear $I-V$ trace. This linear $I-V$'s
in normal magnetic fields and the recovery of the conventional
nonlinear $I-V$ characteristics in tilted fields are the key
features of the topological defects in bilayer systems.

Our samples are from two different wafers, A and B, grown by
molecular beam epitaxy on semi-insulating GaAs substrate. They are
GaAs/AlGaAs quantum wells of different width and shape. For wafer
A, active layers form a 760~\AA\ wide parabolic quantum well,
whereas for wafer B, a 300~\AA\ wide quantum well is of
rectangular form. In the center of each well, a 3 monolayer thick
Al$_x$Ga$_{1-x}$As ($x=0.3$) sheet is grown, which serves as a
tunnel barrier. The symmetrically doped wells are capped by
600~\AA\ Al$_x$Ga$_{1-x}$As ($x=0.3$) and  GaAs layers. Both
structures, A and B,  contain 2DEG of  similar electron
concentration (about $4\times 10^{11}$cm$^{-2}$) and mobility.
They are slightly imbalanced (i.e., have unequal electron
concentrations in two parts of the well).

The samples are patterned in a quasi-Corbino
geometry~\cite{alida}. Ohmic contacts are made to both electron
layers simultaneously. In the experiment we study the $I-V$
characteristics of the gate-gap region at the temperature of 30~mK
in magnetic fields up to 14~T. Tilting the sample plane with
respect to the magnetic field allows us to introduce an in-plane
field component to affect the energy spectrum of the bilayers. The
electron concentration in the ungated region was obtained from
usual magnetoresistance measurements. Also, magnetocapacitance
measurements were performed to determine the energy spectrum under
the gate~\cite{dqw}.

The onset voltage $V_{on}$, at which the system becomes
single-layer, is very different for both samples: $V_{on}=-0.3$~V
for the wafer A and $V_{on}=-0.12$~V for the wafer B. Therefore,
at the filling factor combination $\nu=3, g=2$ which we are
interested in here, the samples are in different isospin
configurations. For sample A, $g = 2$ is obtained at $V_g=-0.19 >
V_{on}$,  so that the sample is in a bilayer regime. Sample B, on
the other hand, is single-layer at $g=2$ ($V_g=-0.15 < V_{on}$).
Both samples are in an equivalent isospin state at the filling
factor combination $\nu=3, g=1$ ($V_g<V_{on}$, resulting in a
single layer under the gate), which can be used as a reference.

\begin{figure}
\includegraphics[width= 0.9\columnwidth]{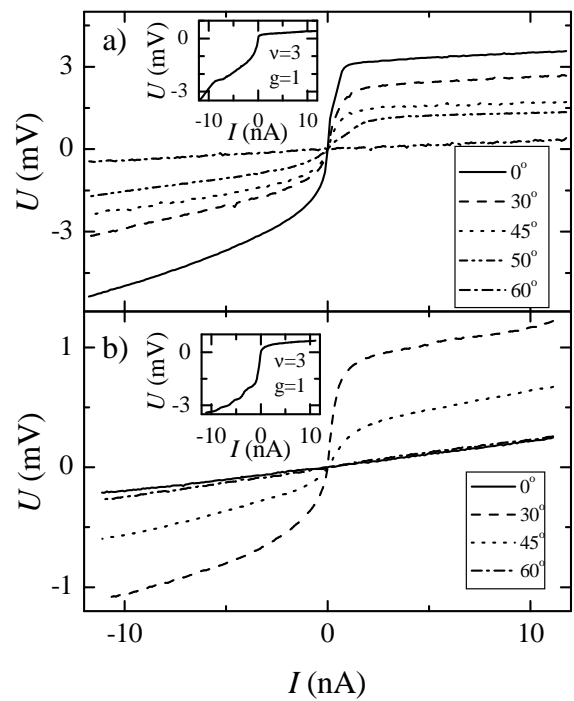}%
\caption{  $I-V$ curves for samples B (a)  and A (b)   for filling
factors $\nu=3$ and $g=2$ at different tilt angles equal to
$\theta=0$ (solid line), $\theta=30^\circ$ (dashed line),
$\theta=45^\circ$ (dotted line), $\theta=50^\circ$ (dash-dot-dot,
only a) panel),$\theta=60^\circ$ (dash-dot line). Perpendicular
magnetic field is constant and is equals to 6.6~T for the sample B
and 5.9~T for the sample A. Inserts demonstrate $I-V$ curves for
samples B and A at filling factors $\nu=3$ and $g=1$ in
perpendicular magnetic field. \label{IV32}}
\end{figure}

In Fig.~\ref{IV32}, $I-V$ curves are presented for both types of
the samples, in normal and tilted magnetic fields. For the sample
that is single-layer at $g=2$ (Fig~\ref{IV32} (a)), the $I-V$
curves are strongly non-linear. They have a well-defined threshold
at positive voltages ($\sim 3$~meV in a perpendicular field),
above which a current starts to flow. The value of the threshold
 corresponds to the subband splitting.

In contrast, the sample with  bilayer  $g=2$ (A-type,
Fig~\ref{IV32} (b)) has a fully linear $I-V$ curve   in
perpendicular  magnetic fields. This demonstrates the absence of a
gap between the inner and outer edge states, in agreement with the
ES picture discussed above (Fig.~\ref{ES} (c), right). Tilting the
sample, while keeping the normal field component constant, causes
the $I-V$ curve to become strongly non-linear. This is in
agreement with the prediction that the in-plane field component
will remove the topological defects and re-establish a gap
structure (see Fig.~\ref{ES} (a), right).

Further increase of the tilt angle flattens $I-V$ curves,
 similarly for both sample types. It is important, that even at
the highest tilt angle $\theta=60^\circ$,  $I-V$'s for both
samples are slightly non-linear, while in a perpendicular field,
the curve is fully linear for the sample of type A.

The conclusion that the different behavior of the two samples
shown in Fig.~\ref{IV32} is related to the different isospin
configuration is further supported by measuring the $I-V$
-characteristics at the filling factor combination $\nu=3, g=1$.
In this case, both samples are in the same isospin configuration,
and the curves are qualitatively similar for both samples in
perpendicular (see inserts to Fig.~\ref{IV32}), as well as in
tilted fields (not shown here). They are non-linear and
asymmetric, with a small positive threshold voltage given by the
spin gap.

As mentioned above, the linear $I-V$ curve of sample A for the
filling factor combination $\nu=3, g=2$ in a perpendicular
magnetic field indicates full equilibration between the inner and
any (or both) of two outer ES  in the gate-gap region.
Figure~\ref{ES} (c), right, suggests that the innermost ES will
only interact with the outermost ES (but not the middle one) if
the spin conservation is assumed. We find that the experimental
$I-V$ slope exactly coincides with one obtained from Buttiker's
calculation~\cite{rdiff} if the equilibration in the gate-gap is
established between two of the ES involved. This allows us to draw
the following conclusions: (i) because the gate-gap width
(2~$\mu$m) is much smaller than the characteristic equilibration
lengths in transport between the ES (of the order of 100$\mu$m,
see Refs.~\onlinecite{haug,mueller}), a defect must be present
which very efficiently couples different ES; (ii) this defect only
couples ES with the same spin in the gate-gap; (iii) the
equilibration by the defect is destroyed by an in-plane field
component. These experimental findings, combined together,  give
strong evidence that the defect is not an impurity of some kind,
but indeed a topological defect induced by the different isospin
configurations in the gated and ungated regions of the sample.

We wish to thank Dr. A.A.~Shashkin and Yu.A.~Melnikov for help
during the experiment. We also thanks Greg Snider for the
Poisson-Schrodinger solver. We gratefully acknowledge financial
support by the RFBR, RAS, the Programme "The State Support of
Leading Scientific Schools", Deutsche Forschungsgemeinschaft, and
SPP "Quantum Hall Systems", under grant LO 705/1-2. E.V.D.
acknowledges support by Russian Science Support Foundation.

\end{document}